\documentclass[12pt]{article}
\usepackage{epsfig}
\usepackage{amsmath}
\usepackage{amssymb}

\def\Journal#1#2#3#4{{#1} {\bf #2}, #3 (#4)}
\def\ANP{{\em Adv. Nucl. Phys.}}
\def\EPJ{{\em Eur. Phys. J.} A}

\def\NPB{{\em Nucl. Phys.} B}

\def\PLB{{\em Phys. Lett.} B}

\def\PRC{{\em Phys. Rev.} C}

\begin{document}

\title{
The Reaction $pp \rightarrow pp\phi$ and the Validity of the OZI Rule
\footnote{Presented at the Baryon98 conference, Bonn, September 1998}}

\author{J. Haidenbauer$^a$, K. Nakayama$^{a,b}$, J. W. Durso$^c$, \\ C. Hanhart$^{a,d}$,
 and J. Speth$^a$ \\ \\
\small{$^a$ Forschungszentrum J\"ulich, IKP, J\"ulich, Germany} \\
\small{$^b$ University of Georgia, Athens, USA} \\
\small{$^c$ Mount Holyoke College, South Hadley, USA} \\
\small{$^d$ ITKP, University of Bonn, Germany}}


\maketitle

\begin{abstract}
The reaction $pp \rightarrow pp \phi$ is investigated within
a relativistic meson-exchange model of hadronic interactions.
The possibility of extracting the $NN\phi$ coupling constant
from a study of this process is explored. 
A combined analysis of $\omega-$ and $\phi$-meson production
in proton-proton collisions utilizing recent data on these
reactions is carried out and yields values for $g_{NN\phi}$ that 
are compatible with the OZI rule. 
\end{abstract}

\section{Introduction}

Recent experiments on $\bar pp$ annihilation at rest 
revealed unexpectedly large cross section 
ratios $\sigma_{\bar pp \rightarrow \phi X} / 
\sigma_{\bar pp \rightarrow \omega X}$, exceeding the estimate from the 
OZI rule by one order of magnitude or even more 
(cf. Ref.\cite{Ell95} for a compilation of data). 
These large $\phi$-production cross sections were interpreted by some 
authors as a clear signal for an intrinsic 
$\bar ss$ component in the nucleon\cite{Ell95}. However, in 
an alternative approach based on two-step processes these data
can be explained without introducing any strangeness 
in the nucleon and any explicit violation of the OZI rule\cite{Loche}.

In this context $\phi$ production in nucleon-nucleon collisions is
of specific interest. Here one does not expect any significant
contributions from competing OZI-allowed two-step mechanisms.
Therefore cross section ratios 
$\sigma_{pp\rightarrow pp \phi} / \sigma_{pp\rightarrow pp \omega}$
should provide a clear indication for a possible OZI violation and 
the amount of hidden strangeness in the nucleon. Indeed preliminary
data presented recently by the DISTO collaboration indicate that this
ratio is about 8 times larger than the OZI estimate\cite{Disto}. 

In this work 
we report on a model analysis of the DISTO data with the aim of extracting
the $NN\phi$ coupling constant. Specifically we want to see whether the
observed enhancement over the OZI estimate in the cross section implies
a $g_{NN\phi}$ that is likewise enhanced and therefore 
at variance with the OZI rule. 

\section{The Model}

We describe the $pp \rightarrow pp\phi$ reaction within a 
relativistic meson-exchange model, where 
the transition amplitude is calculated in Distorted Wave Born Approximation  
in order to take the $NN$ final state interaction into account. (See 
Ref.\cite{Nak1} for the details of the formalism.) 
For the $NN$ interaction we employ the model Bonn B\cite{MHE87}. 
We do not consider the initial state interaction explicitly.  
Its effect is accounted for via an appropriate adjustment of the 
(phenomenological) form factors at the hadronic vertices.

\begin{figure}[ht]
\begin{center}
\epsfig{figure=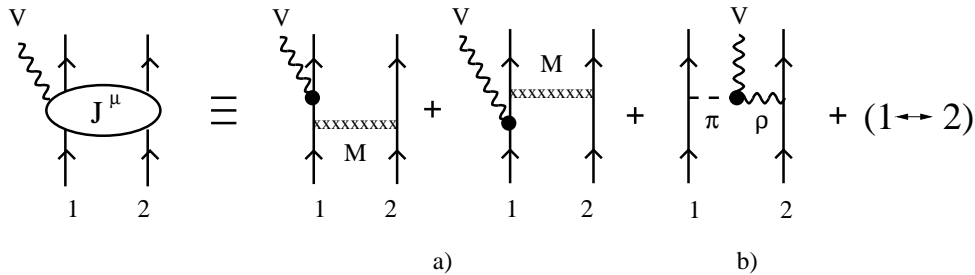,height=3.5cm}
\caption{$\omega$ and $\phi$ meson production mechanisms: a) nucleonic
current; b) $V\rho\pi$ meson-exchange current.}
\label{fi1}
\end{center}
\end{figure}

In a previous study of the reaction $pp\rightarrow pp\omega$\cite{Nak1}
we found that the dominant production mechanisms are the nucleonic and 
$\omega\rho\pi$ mesonic currents, as depicted in Fig.~\ref{fi1}. 
We also found that the angular distribution of the produced $\omega$
meson provides a unique and clear signature of the magnitude of these
currents, thus allowing one to disentangle these two reaction mechanisms.
The situation is quite similar for the reaction $pp\rightarrow pp\phi$. 
In this case the nucleonic
current and the $\phi\rho\pi$-exchange current provide the dominant 
contributions to the production amplitude\cite{Nak2}. 
Therefore, it is possible
to fix uniquely the magnitudes of the nucleonic and the meson-exchange 
current by analyzing the angular
distribution of the $\phi$ meson measured by the DISTO 
collaboration\cite{Disto}.
Furthermore, since the $NN\phi$ coupling constant enters only in the 
nucleonic current it is possible to extract its value from such an
analysis. It is determined by the requirement of getting the 
proper contribution of the nucleonic current needed to reproduce the 
angular distributions.  
 
The parameters of the model (coupling constants, cutoff masses of the
vertex form factors)
are mostly taken over from the employed $NN$ model. The $\phi\rho\pi$ 
coupling constant is obtained from the measured decay width of 
$\phi \rightarrow \rho + \pi$. However, besides the $g_{NN\phi}$ that we 
want to 
extract from the analysis, there are still some more free parameters:
The cutoff mass of the $\phi\rho\pi$ vertex form factor
of the meson-exchange current, and the form factor and tensor- to 
vector coupling constant ratio $\kappa_\phi \equiv f_{NN\phi}/g_{NN\phi}$ 
of the nucleonic current (cf. Ref.\cite{Nak2} for details). 
Thus in order to determine these remaining four parameters 
we need at least four independent $\phi$-meson production data. 
In view of the more limited data set presently available,
we decided to perform a combined analysis of $\phi$- and $\omega$-meson 
production. This means that we use the data from both reactions and we
assume relations between corresponding parameters in the production
amplitudes.
Specifically, we assume that the form factors at the $\phi\rho\pi$-
and $\omega\rho\pi$ vertices are the same. This is a reasonable choice 
because the off-shell particles ($\rho$, $\pi$) are the same in both cases. 
Likewise
we assume that the form factor at the meson production vertices in the
nucleonic current are the same. The $NN\omega$ coupling constant is fixed
to the SU(3) value, $g_{NN\omega}$ = 9, based on $g_{NN\rho}$ of 
Ref.\cite{rhocoup}. Furthermore we assume that $\kappa_\phi = 
\kappa_\omega$, as also suggested by SU(3) symmetry,
and $-0.5 \leq \kappa_\omega \leq 0.5$. 

Concerning the data we use the total cross sections of Ref.\cite{Saclay} 
for the reaction $pp\rightarrow pp\omega$ and the 
$\phi$-meson angular distribution at $T_{lab} = 2.85$ GeV measured
by the DISTO collaboration\cite{Disto}.  
The latter is given without absolute normalization in Ref.\cite{Disto}.
But since the DISTO group has also measured the ratio 
$\sigma_{pp\rightarrow pp\phi} / \sigma_{pp\rightarrow pp\omega}$
at this energy we can estimate $\sigma_{pp\rightarrow pp\phi}$ 
by multiplying this ratio with the total cross section for 
$\omega$-meson production interpolated from the existing data. This 
yields a value of $\sigma_{pp\rightarrow pp\phi} \approx 0.3 \ \mu b$.

\section{Results and Discussion}

\begin{figure}[ht]
\begin{center}
\epsfig{figure=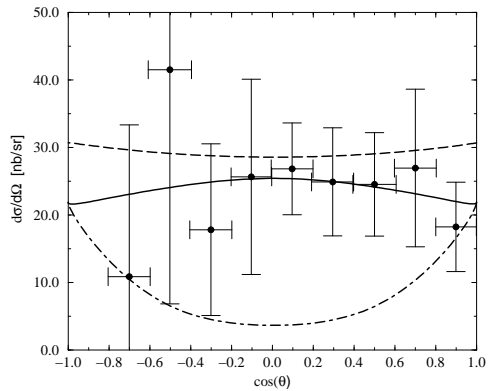,height=6cm}
\caption{Angular distribution of the $\phi$ meson at $T_{lab}=2.85$ GeV.
The dashed (dash-dotted) curve is the result with the mesonic (nucleonic)
current alone. The solid curve is the total result. The data points are
from Ref.\protect\cite{Disto}, normalized as discussed in the text.}
\label{fi2}
\end{center}
\end{figure}

After the above considerations, we are now prepared to apply the model 
to the reactions $pp\rightarrow pp\omega$ and $pp\rightarrow pp\phi$.
The angular distribution for $\phi$-meson production measured at 
$T_{lab} = 2.85$ GeV is shown in Fig.~\ref{fi2}. 
We observe that the angular distribution is fairly flat. Recalling the 
results we obtained for $\omega$ production\cite{Nak1} this 
tells us that $\phi$-meson production should be almost entirely due to the 
$\phi\rho\pi$ meson-exchange current. Only a very small contribution 
of the nucleonic current is required if the angular 
distribution drops at forward and backward angles, as indicated by
the data. 

Our strategy for fixing the various parameters is outlined in detail
in Ref.\cite{Nak2}. Here we only want to mention that the lack of a more 
complete set of data prevents us from achieving a unique determination 
of $g_{NN\phi}$. Rather we get a set of values 
which range from $g_{NN\phi}=-0.163$ to $g_{NN\phi}=-1.40$.
Nevertheless, it is encouraging to see that the extracted values all lie within
fairly narrow bounds. This clearly indicates to us that the dependence on
the model parameters is not very strong, and that the magnitude of 
$g_{NN\phi}$ is primarily determined by the experimental information used. 

The values of $g_{NN\phi}$ obtained may be compared with those
resulting from SU(3) flavor symmetry considerations and imposition of the 
OZI rule,  
$$
g_{NN\phi} = - 3 g_{NN\rho} \sin(\alpha_v) \cong  -(0.60 \pm 0.15) \ ,
$$
where the factor $\sin(\alpha_v)$ is due to the deviation from the ideal
$\omega - \phi$ mixing. The numerical value is obtained using the values 
of $g_{NN\rho} = 2.63 - 3.36$\cite{rhocoup} and $\alpha_v \cong 3.8^o$. 
Comparing this value with the ones extracted from our model analysis, we 
conclude that the preliminary data presently available can be described with 
using a $NN\phi$ coupling constant that is compatible with the OZI rule.
This clearly indicates that a dynamical model is needed for drawing
any conclusion about the validity of the OZI rule.


\end{document}